%% file: LIMoSim_OMNeT.tex
\documentclass[%
	conference,
	a4paper,
	10pt
]{IEEEtran}

\usepackage{graphicx}
\usepackage{epstopdf}
\usepackage{standalone}
\usepackage[nolist ]{acronym}

\usepackage{xstring}

\input{tex/packages.tex}

\renewcommand{\baselinestretch}{0.91}

\begin{document}

\input{tex/variables.tex}

\input{tex/acro.tex}

\acresetall

\input{tex/title}

\input{tex/abstract.tex}

\IEEEpeerreviewmaketitle

\input{tex/introduction.tex}

\input{tex/relatedWork.tex}

\input{tex/solution_approach.tex}

\input{tex/results.tex}

\input{tex/conclusion.tex}

\input{tex/acknowledgment.tex}

\bibliographystyle{IEEEtran}


\end{document}

%% file: tex/packages.tex
\usepackage[utf8]{inputenc}
\usepackage{marvosym}

\usepackage{listings,lstautogobble}
\usepackage{algorithm}
\usepackage{algpseudocode}
\usepackage{tikz}

\usepackage{pgfplots}
\pgfplotsset{compat=newest}
\pgfplotsset{
    box plot/.style={
        /pgfplots/.cd,
        fill=blue!30,
        only marks,
        mark=-,
        mark size=0.2em,
        /pgfplots/error bars/.cd,
        y dir=plus,
        y explicit,
    },
    box plot box/.style={
        /pgfplots/error bars/draw error bar/.code 2 args={%
            \draw  ##1 -- ++(.2em,0pt) |- ##2 -- ++(-.2em,0pt) |- ##1 -- cycle;
        },
        /pgfplots/table/.cd,
        y index=2,
        y error expr={\thisrowno{3}-\thisrowno{2}},
        /pgfplots/box plot
    },
    box plot top whisker/.style={
        /pgfplots/error bars/draw error bar/.code 2 args={%
            \pgfkeysgetvalue{/pgfplots/error bars/error mark}%
            {\pgfplotserrorbarsmark}%
            \pgfkeysgetvalue{/pgfplots/error bars/error mark options}%
            {\pgfplotserrorbarsmarkopts}%
            \path ##1 -- ##2;
        },
        /pgfplots/table/.cd,
        y index=4,
        y error expr={\thisrowno{2}-\thisrowno{4}},
        /pgfplots/box plot
    },
    box plot bottom whisker/.style={
        /pgfplots/error bars/draw error bar/.code 2 args={%
            \pgfkeysgetvalue{/pgfplots/error bars/error mark}%
            {\pgfplotserrorbarsmark}%
            \pgfkeysgetvalue{/pgfplots/error bars/error mark options}%
            {\pgfplotserrorbarsmarkopts}%
            \path ##1 -- ##2;
        },
        /pgfplots/table/.cd,
        y index=5,
        y error expr={\thisrowno{3}-\thisrowno{5}},
        /pgfplots/box plot
    },
    box plot median/.style={
        /pgfplots/box plot
    },
    boxplot/every median/.style={
    	ultra thick,dashed,cyan
    }
}

\definecolor{flexicolor}{RGB}{46,49,146}
\definecolor{amaricolor}{RGB}{237,28,36}

\usepackage{xspace}

\usepackage[binary-units=true]{siunitx}
\usepackage{psfrag}
\usepackage{graphicx}
\usepackage{tabularx,booktabs}
\usepackage{multirow}
\usepackage{rotating}

\renewcommand{\baselinestretch}{1}

\usepackage{multirow}

\usepackage[cmex10]{amsmath}
\usepackage[caption=false,font=footnotesize]{subfig}
\usepackage{stfloats}

%% file: tex/variables.tex
\newcommand{\paperTitle}{LIMoSim: A Lightweight and Integrated Approach for Simulating Vehicular Mobility with OMNeT++}
\newcommand{\paperAuthors}{Benjamin Sliwa, Johannes Pillmann, Fabian Eckermann and Christian Wietfeld}
\newcommand{\paperEmails}{$\{$Benjamin.Sliwa, Johannes.Pillmann, Fabian.Eckermann, Christian.Wietfeld$\}$@tu-dortmund.de}

\newcommand{\githubUrl}{\footnote{Available at https://github.com/BenSliwa/LIMoSim} }

\newcommand{\figurePadding}{0pt}
\newcommand{\figureTopPadding}{\figurePadding}
\newcommand{\figureBottomPadding}{\figurePadding}

%% file: tex/acro.tex
\begin{acronym}
	\acro{OSM}{OpenStreetMap}
	\acro{VACaMobil}{VANET Car Mobility Manager}
	
	\acro{IPC}{Interprocess Communication}
	\acro{TraCI}{Traffic Control Interface}
	
	\acro{IDM}{Intelligent Driver Model}
	\acro{MOBIL}{Minimizing Overall Braking Induced by Lane change}
	
	\acro{SWIM}{Small Worlds In Motion}
	
	\acro{DES}{Discrete Event Simulation}
	
	\acro{UI}{User Interface}
	\acro{RSSI}{Received Signal Strength Indicator}
	
	\acro{HO}{Handover}
	
	\acro{CAT}{Channel-aware Transmission}
	\acro{pCAT}{predictive CAT}
	\acro{MTC}{Machine-Type Communication}
	
	\acro{V2V}{Vehicle-to-Vehicle}
	\acro{V2I}{Vehicle-to-Infrastructure}

	\acro{LTE}{Long Term Evolution}
	\acro{LTE D2D}{LTE Device-to-Device}
	\acro{UE}{User Equipment}
	\acro{eNB}{Evolved Node B}
	\acro{IoT}{Internet of Things}

	\acro{AODV}{Ad hoc On-demand Distance Vector}
	\acro{B.A.T.M.A.N.}{Better Approach To Mobile Adhoc Networking}
	\acro{OLSR}{Optimized Link State Routing}

	\acro{OFDM}{Orthogonal Frequency-Division Multiple Access}

	\acro{KPI}{Key Performance Indicator}
	\acro{PDR}{Packet Delivery Ratio}

	\acro{MANET}{Mobile Ad-hoc Network}
	\acro{VANET}{Vehicular Ad-hoc Network}
	\acro{WMN}{Wireless Mesh Network}
	\acro{WSN}{Wireless Sensor Network}

	\acro{RREP}{Route Reply}
	\acro{RREQ}{Route Request}

	\acro{OMNeT++}{Objective Modular Network Testbed in C++}
	\acro{SUMO}{Simulation of Urban Mobility}
	\acro{Veins}{Vehicles in Network Simulation}
	\acro{NS2}{Network Simulator 2}
	\acro{NS3}{Network Simulator 3}
	\acro{LIMoSim}{Lightweight ICT-centric Mobility Simulation}

	\acro{IPv4}{Internet Protocol version 4}

	\acro{QoS}{Quality of Service}
	\acro{TOS}{Type of Service}
	
	\acro{MAC}{Medium Access Control}

	\acro{UDP}{User Datagram Protocol}
	\acro{TCP}{Transmission Control Protocol}
	\acro{MTU}{Maximum Transmission Layer}

	\acro{UAV}{Unmanned Aerial Vehicle}
	\acro{UGV}{Unmanned Ground Vehicle}
	
\end{acronym}

%% file: tex/title.tex
\title{\paperTitle}

\author{\IEEEauthorblockN{\paperAuthors}
	\IEEEauthorblockA{Communication Networks Institute\\
		TU Dortmund\\
		44227 Dortmund, Germany\\ e-mail: \paperEmails}}
\maketitle

%% file: tex/abstract.tex
\begin{abstract}
	
%
%
Reliable and efficient communication is one of the key requirements for the deployment of self-driving cars. Consequently, researchers and developers require efficient and precise tools for the parallel development of vehicular mobility and communication.
%
%
Although current state-of-the-art approaches allow the coupled simulation of those two components, they are making use of multiple specialized simulators that are synchronized using interprocess communication, resulting in highly complex simulation setups. Furthermore, the compatibility of those simulators requires constant attention as they are developed independently.
%
%
In this paper, we present a lightweight and integrated approach for simulating vehicular mobility directly in \ac{OMNeT++} and INET without the need for external tools or \ac{IPC}. The proposed framework \ac{LIMoSim} is available as Open Source software and can easily be combined with other third-party extension frameworks for providing vehicular mobility based on well-known microscopical models. In contrast to existing approaches, the amount of necessary preprocessing steps for simulation setups is significantly reduced.
%
%
The capabilities of \ac{LIMoSim} are demonstrated by a proof of concept evaluation in combination with the \ac{LTE} simulation framework SimuLTE.

\end{abstract}

%% file: tex/introduction.tex
\section{Introduction}

\begin{figure}[b!]  
	\vspace{\figureTopPadding}
	\centering		  
	\includegraphics[width=1\columnwidth]{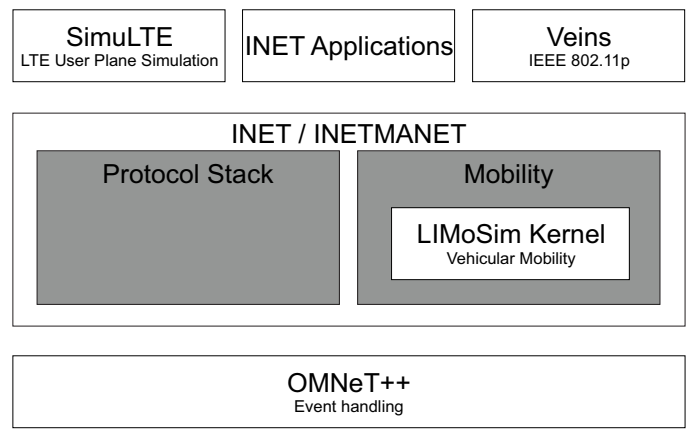}
	\caption{Integration of the proposed framework into the family of \ac{OMNeT++} extensions. Since the the kernel of \ac{LIMoSim} is embedded into the INET mobility module, the integration is transparent for the application-specific extensions frameworks like SimuLTE and can be used without requiring additional adjustments.}
	\label{fig:scenario}
	\vspace{\figureBottomPadding}	
\end{figure}

%
%
\ac{OMNeT++} \cite{VargaHornig2008} is a well-established network simulation framework available as Open Source software for academic usage. Because of its modular approach, it has been extended by many third-party frameworks focusing on specialized communication technologies like \ac{LTE} and IEEE 802.11p.
%
%
As we have shown in previous work \cite{SliwaBehnkeIdeEtAl2016} the performance and robustness of communication systems can be highly improved by integrating knowledge of the users' mobility behavior into the routing decisions. With the deployment of autonomous cars in the near future, information about planned trajectories becomes available in all those vehicles and should be exploited for decision processes in the next generation of intelligent communication systems.
With the increasing amount of interactions between mobility and communication, engineers and developers require efficient tools for simulating both aspects at once. Although current state-of-the-art frameworks already allow the coupling of specialized simulators for the two individual aspects, this approach has a number of system-immanent disadvantages (cf. Sec.~\ref{sec:relatedWork}) because each of those simulators was developed for a very isolated field of application and not intended to be combined with others.
%
%
In this paper, we present a lightweight framework for simulating microscopic vehicular traffic based on well-known models. In contrast to existing approaches, the proposed \ac{LIMoSim} is intended to be used by communication simulators by design. It is seamlessly integrated into \ac{OMNeT++} (cf. Fig.~\ref{fig:scenario}) and requires no external tools or synchronization through \ac{IPC}. Furthermore, it is completely compatible to third-party extension frameworks like SimuLTE \cite{VirdisSteaNardini2015} and INETMANET, which can integrate the novel mobility modules into their simulation scenarios in a transparent way using the widely-used INET framework.
%
%
The rest of the paper is structured as follows. After discussing the related work, we present the basic architecture of our simulation model and provide a detailed description about the integration of \ac{LIMoSim} into \ac{OMNeT++}. In the next section, we describe the simulation setup for a proof of concept evaluation scenario in an \ac{LTE}-context. Finally, detailed simulation results are presented and discussed.

%% file: tex/relatedWork.tex
\section{Related Work} \label{sec:relatedWork}
\ac{Veins} \cite{SommerGermanDressler2011} is a well-established simulation framework for simulating \acp{VANET} in \ac{OMNeT++}. It provides an implementation of IEEE 802.11p and acts as an interface to the microscopic traffic simulator \ac{SUMO} \cite{KrajzewiczErdmannBehrischEtAl2012}. Both simulators are synchronized through \ac{IPC} using \ac{TCP} and the dedicated \ac{TraCI} protocol of \ac{SUMO}.
While this approach ensures highly precise simulation results for both communication and vehicular mobility through usage of specialized simulators, it has a number of disadvantages:
\begin{itemize}
	\item Since the different simulators are developed individually, their \emph{compatibility} needs to be validated with every new release. When we started to work on \ac{LIMoSim}, \ac{Veins} was not compatible to the newest \ac{SUMO} version. 
	\item For mobility-aware communication applications, \ac{TraCI} is the  bottleneck for the development process. The protocol needs to be extended for every new information type that should be shared between the simulators.
	\item Simulation setups have a high complexity because different tools have to be executed simultaneously. This aspect is even more dramatic for massive simulation scenarios that are executed on multiple servers in parallel. Additionally, even the generation of \ac{SUMO}-only scenarios from \ac{OSM}-data is quite complex, as several preprocessing steps are required.
	\item \ac{SUMO} itself is rather designed for being used in a static way using precomputed data (for example routing paths). This does not match well with highly-dynamic vehicular applications, where the mobility behavior influences communication processes and vice versa.
\end{itemize}
Since the \ac{IPC}-approach using \ac{SUMO} is the current state-of-the-art way for simulating vehicular mobility in \ac{OMNeT++}, those disadvantages are propagated to third-party extension frameworks like INETMANET and SimuLTE as well, if they want to make use of vehicular motion.

%
%
Consequently, we think the \ac{OMNeT++} community could benefit from a more diverse way for simulating vehicular mobility depending on the application scenario. While the current approach using  \ac{Veins} with \ac{SUMO} is fitting for many IEEE 802.11p applications, there are scenarios where \ac{Veins} could be coupled with our proposed framework to avoid the \ac{SUMO} overhead and the need for \ac{IPC}. Additionally, the complexity of many \ac{LTE} and \ac{MANET} scenarios can be significantly reduced by using \ac{LIMoSim} only.

%% file: tex/solution_approach.tex
\section{Integration of \ac{LIMoSim} into OMNeT++}

\ac{LIMoSim} focuses on highly dynamic traffic scenarios where all decision processes and routes are determined at runtime. Its main field of application is the simulation of medium-sized city scenarios, where intelligent vehicles interact with other traffic participants through means of communication. In this context, the simulation of vehicular mobility is considered as a service for the simulation of communication systems. 

The simulator consists of two main components: the simulation kernel with the different elements of the microscopic mobility models and the \ac{UI} part for standalone vehicle simulation and the road editor for easy generation of new mobility scenarios. 
It furthermore features live visualization of statistical data in dynamically updated plots and contains export functions for vector graphics.
In this paper, we focus on the simulation kernel of \ac{LIMoSim}, as it is the only component that is linked to \ac{OMNeT++} and provide descriptions of the hierarchical mobility model and the event handling mechanism in the following subchapters.

\subsection{Simulation of Vehicular Mobility}

\begin{figure}[t]  	
	\vspace{\figureTopPadding}
	\centering		  
	\includegraphics[width=1\columnwidth]{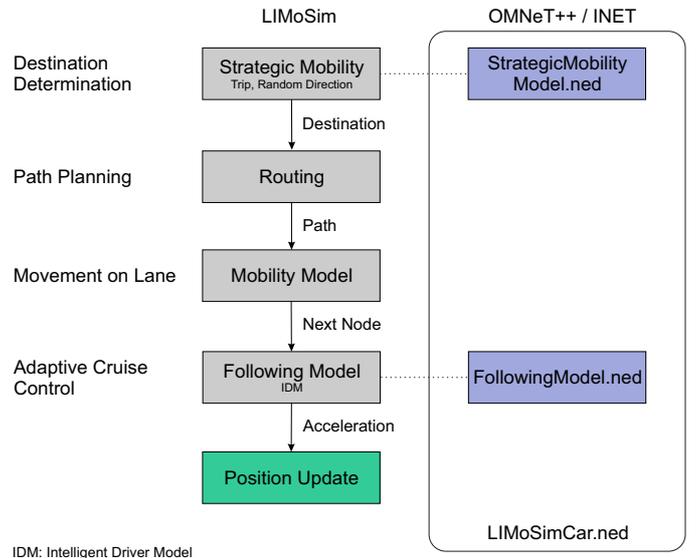}
	\caption{Hierarchical model for microscopic traffic simulation encapsulated in the \emph{LIMoSimCar.ned} module. \ac{OMNeT++}-interface modules are available for the top and the bottom layer of the model.}
	\label{fig:approach}
	\vspace{\figureBottomPadding}	
\end{figure}

The \emph{LIMoSimCar.ned} module extends the \emph{MovingMobilityBase.ned} module of the INET framework and acts as a logical structure for the different mobility-related submodules. Its hierarchical structure is illustrated in Fig.~\ref{fig:approach}.
A strategic model is used to determine the current destination point, which is the motivation for the vehicles movement. This node is then passed to a routing algorithm that computes the optimal path with respect to a defined criterion. All nodes of the path are handled sequentially by the general mobility model. Required adjustments to the mobility behavior because of encounters with other traffic participants are handled by the \emph{Following Model}, which controls the distance between vehicles depending on their velocity by acceleration and deceleration. Finally, the position is updated with the calculated speed. In the current version of \ac{LIMoSim}, we use the \ac{IDM} \cite{TreiberHenneckeHelbing2000} as a well-known following model for providing a realistic representation of vehicular mobility and combine it with the dedicated \emph{lane-changing model} \ac{MOBIL} \cite{KestingTreiberHelbing2007}. It should be noted that since the regular version of this model is not intersection-aware, the IDM implementation of LIMoSim treats traffic signals like static vehicles if the light is yellow or red. The integration of further models is planned for later releases. 
\begin{figure}[h]
	\begin{lstlisting}
	*.ue.mobilityType = "LIMoSimCar"
	*.ue.mobility.map = "map.osm"
	*.ue.mobility.strategicModel = "Trip"
	*.ue.mobility.strategicModel.trip = "677230875,
	275672221,3569208993,477807"
	*.ue.mobility.way = "337055293"
	*.ue.mobility.segment = 4
	*.ue.mobility.lane = 0
	*.ue.mobility.offset = 1m

	\end{lstlisting}
	\caption{Example .ini configuration for using \ac{LIMoSim} in an \ac{LTE}-scenario. The trip model routes the car to a sequential list of a destinations. The actual node ids can be obtained from the \ac{UI} part of \ac{LIMoSim} or the raw \ac{OSM}-files. The configuration can be either done manually by setting the parameters of the mobility modules or automatically using a generated XML file.}
	\label{fig:mobilityConfiguration}
\end{figure}
Since the individual mobility control modules are interfaced through a \emph{Simple Module}, the respective parameters can be set for the \emph{.ned-file} without requiring external configuration files. An example configuration is shown in Fig.~\ref{fig:mobilityConfiguration}, where a \ac{UE} is assigned with the \texttt{LIMoSimCar} mobility type and positioned on a specified way segment. Furthermore, the \texttt{Trip} strategic mobility model is configured to sequentially approach multiple destination points (cf. Fig.~\ref{fig:simulation}). The ids originate from the \ac{OSM} data model and can be obtained using \ac{UI} of \ac{LIMoSim}. 

\subsection{Embedding \ac{LIMoSim}-Events into OMNeT++}

\begin{figure}[h]  	
	\vspace{\figureTopPadding}
	\centering		  
	\includegraphics[width=1\columnwidth]{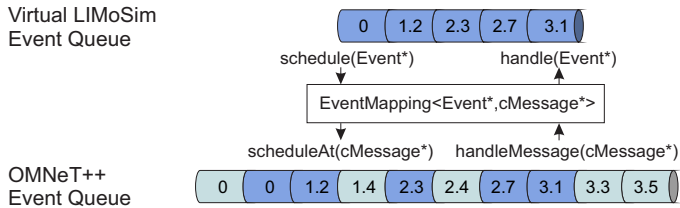}
	\caption{Coupling of the event queues for mobility and communication simulation. \ac{LIMoSim} objects are not aware of their execution environment and use a virtual event queue, which is mapped to the event queue of \ac{OMNeT++}.}
	\label{fig:des}
	\vspace{\figureBottomPadding}	
\end{figure}

The \ac{DES}-coupling mechanism is illustrated in Fig.~\ref{fig:des}. Since \ac{LIMoSim} can also be used in a standalone mode, its objects are not aware of their \ac{OMNeT++}-environment. The integration into the event handling mechanism is performed by a virtual event queue that does not take \ac{OMNeT++} events into consideration. In contrast to the \ac{IPC}-based approach, there is no need for real \ac{DES}-synchronization, as only a single queue is used. If an event $e$ is scheduled, the \emph{event mapping singleton} creates a new \ac{OMNeT++} \emph{cMessage} $m$ and stores a map entry for original event. $m$ is then inserted into the \ac{OMNeT++} event queue. Once the \emph{handleMessage()} method for $m$ is called, $e$ is retrieved from the map entry and handled by the respective object.
With this approach, the actual event handling in \ac{OMNeT++} is transparent for all \ac{LIMoSim} objects. Moreover, \ac{LIMoSim} objects can use the event handling mechanism of \ac{OMNeT++} without even requiring an actual \ac{OMNeT++} module. As a consequence, objects that only influence the vehicular traffic but do not contain communication modules (like interference traffic or traffic signals) do not need to be modelled in \ac{OMNeT++}.

%% file: tex/results.tex
\section{Proof of Concept Evaluation}

In this section, we present the setup and the results of the proof of concept evaluation. The default parameters for the scenarios are defined in Tab.~\ref{tab:simulation_parameters}. Additional \ac{LTE}-parameters are set according to the \ac{HO} example of SimuLTE.
\input{tex/simulation_parameters.tex}

\subsection{LTE-scenario with Real-world Map Data} \label{chap:lte_evaluation}

\begin{figure}[h]  	
	\vspace{\figureTopPadding}
	\centering		  
	\includegraphics[width=1\columnwidth]{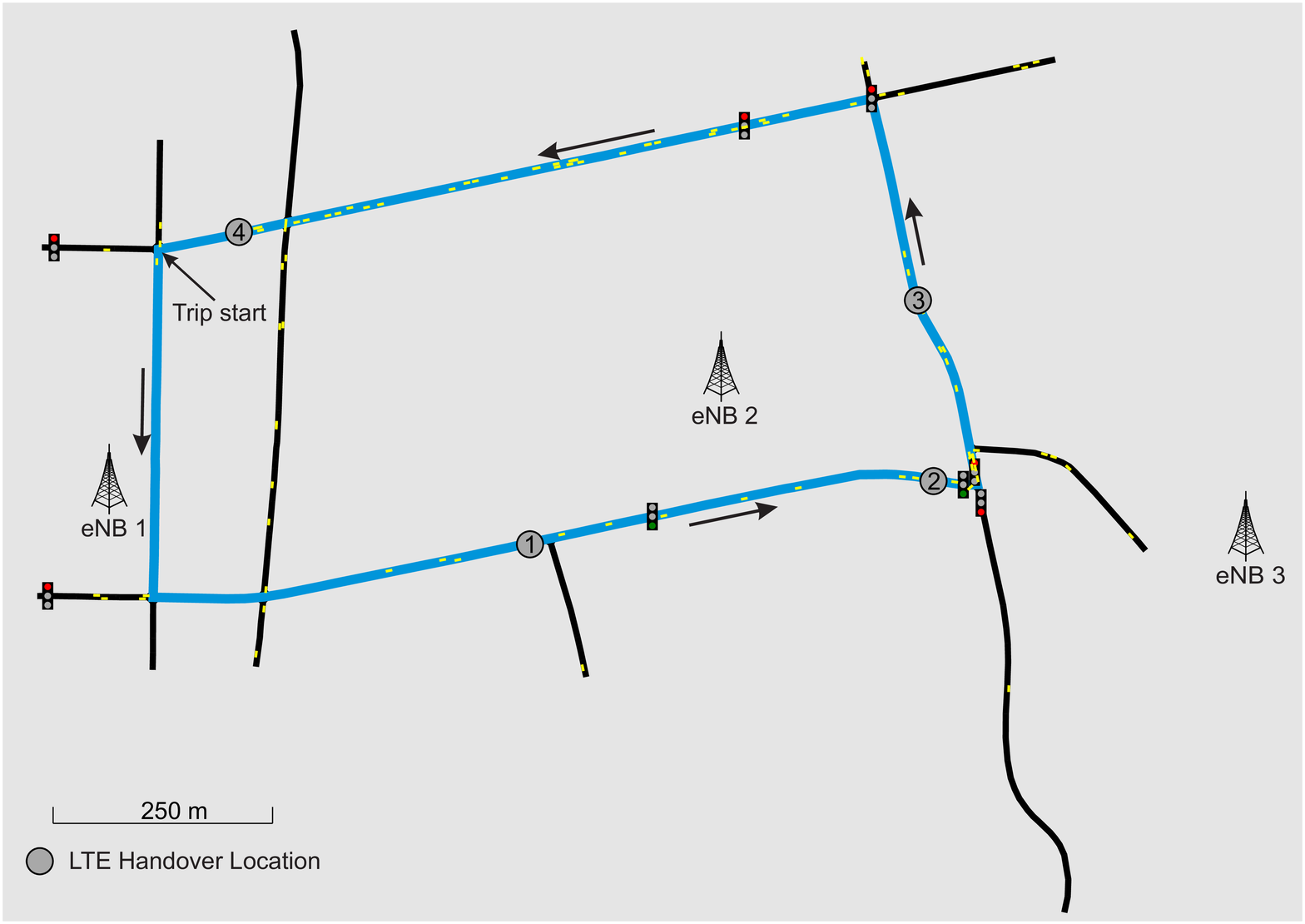}
	\caption{Example scenario using \ac{OSM} location data. 100 cars are simulated in the TU Dortmund University area. The map has been exported directly from the UI part of \ac{LIMoSim} using its vector graphic export function.}
	\label{fig:simulation}
	\vspace{\figureBottomPadding}	
\end{figure}

\begin{figure}[t]  	
	\vspace{\figureTopPadding}
	\centering		  
	\includegraphics[width=1\columnwidth]{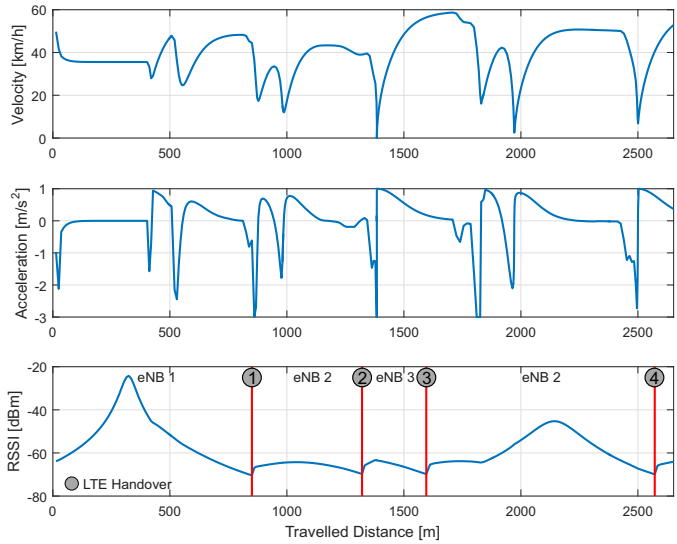}
	\caption{Temporal behavior for velocity, acceleration and \ac{RSSI} of the considered vehicle. The graphs show typical inner-city mobility characteristics and plausible \ac{LTE}-behavior.}
	\label{fig:evaluation}
	\vspace{\figureBottomPadding}	
\end{figure}

As a realistic scenario, real-world map data from \ac{OSM} is used. An \ac{LTE}-enabled car is monitored while it is driving around the campus area of the TU Dortmund University using a \emph{Trip} strategic mobility module. 100 other cars act as interference traffic with \emph{Random Direction} mobility and influence the mobility behavior of the considered car. The area is covered by three different \acp{eNB}, which have been positioned according to network provider information. Fig.~\ref{fig:simulation} provides an illustration of the described simulation setup. Due to the mobility of the car, multiple \acp{HO} are required at runtime.
For the considered car, the current velocity, acceleration and the \ac{RSSI} of the \ac{LTE} link are measured over the trip duration and shown in Fig.~\ref{fig:evaluation}.

%
%
The graphs for velocity and acceleration show typical inner-city characteristics and mirror the traffic dynamics of the scenario. Due to the map topology, the traffic signals and the other traffic participants, the current street condition is under constant change. Therefore, a lot of braking and accelerating is required, causing a highly dynamic velocity behavior. 
%
%
The \ac{RSSI} of the \ac{LTE} signal shows plausible characteristics depending on the distance to the current serving \ac{eNB}. Four handovers occur during the considered time (for the actual handover locations cf. Fig.~\ref{fig:simulation}). With \ac{HO}2, the device attaches from \ac{eNB}2 to \ac{eNB}3 and  shortly afterward back to \ac{eNB}2 with \ac{HO}3. This \emph{ping pong} behavior can be considered as a motivation for further research in the interdependency of communication and mobility, as the handovers could have probably been avoided using a mobility-aware decision approach.
In earlier work \cite{PillmannSliwaSchmutzlerEtAl2017}, we have performed a similar case-study using a different toolchain with \ac{SUMO} and an analytical \ac{LTE} model, which achieved similar results for the handover behavior.

\subsection{Example Behavior of the IDM model}

\begin{figure}[b]  	
	\vspace{\figureTopPadding}
	\centering		  
	\includegraphics[width=1\columnwidth]{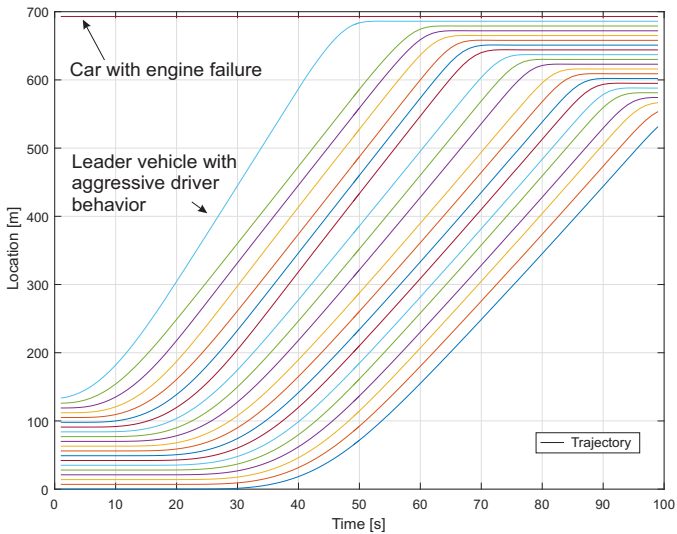}
	\caption{Space-Time diagram for an inner-city scenario: the vehicles start in a jam situation, move into free traffic and are stopped by an obstacle in about 700m.}
	\label{fig:space_time}
	\vspace{\figureBottomPadding}	
\end{figure}

In order to analyze the effects of the \ac{IDM} model on the traffic dynamics, we consider a typical street scenario. Fig.~\ref{fig:space_time} visualizes the acceleration behavior of multiple cars in a space-time diagram. The cars start in a jam situation, which is then resolved into free traffic. After about 700m the cars encounter a car with engine failure that causes another traffic jam as only a single lane is used. Different driver behavior types can be identified depending on the slope of the curve. The results are confirmed by the analytical evaluation in \cite{TreiberKesting2013}.

%% file: tex/simulation_parameters.tex
\begin{table}[h]
	\centering
	\caption{Simulation Parameters of the Reference Scenario}
	\begin{tabularx}{0.44\textwidth}{|X|X|}
		\hline
		
		\textbf{Simulation parameter} & \textbf{Value} \\ \hline

		Strategic mobility model (\ac{UE}) & Trip \\ \hline
		Strategic mobility model (interference traffic) & Random Direction \\ \hline
		Number of interference cars & 100 \\ \hline
		Following model & IDM  \\ \hline
		Lane change model & MOBIL  \\ \hline		
	    Speed factor (driver behavior) & $1 \pm 0.2$ \\ \hline

		Carrier frequency &  1800 [MHz] \\ \hline
		eNode B transmission power & 46 [dBm]  \\ \hline
		eNode B antenna & omnidirectional \\ \hline
		
	\end{tabularx}
	\label{tab:simulation_parameters}
\end{table}

%% file: tex/conclusion.tex
\section{Conclusion}

%
%
In this paper, we presented the novel framework LIMoSim\githubUrl for simulating microscopic vehicular mobility directly in \ac{OMNeT++}.
%
%
In contrast to existing approaches that treat vehicular mobility and communication separately and require \ac{IPC} for the synchronization of different simulation tools, our proposal brings these aspects together in an integrated way.
%
%
The actual mobility simulation relies on well-known models in order to guarantee the required accuracy. The easy integration is especially attractive for \ac{LTE} and \ac{MANET} simulations, which can make use of vehicular mobility without requiring a complex simulation setup.
%
%
The capabilities of \ac{LIMoSim} were demonstrated with a proof-of-concept evaluation in an \ac{LTE}-context using real-world map data from \ac{OSM}.
%
%

In future work, we want to couple \ac{LIMoSim} with \ac{Veins} for the simulation of IEEE 802.11p networks. \ac{LIMoSim} could serve as an alternative to \ac{SUMO} providing a lightweight solution for simulating vehicular motion without the \ac{IPC}-overhead.
Furthermore, we want to integrate strategic mobility models that are closer to human decision making like \ac{SWIM} \cite{MeiStefa2009}, which has already been applied to \ac{OMNeT++} in \cite{UdugamaKhalilovMuslimEtAl2016} and utilize the framework for the simulation of indoor robotic networks in a logistical context. 
Moreover, we want to exploit the visualization capabilities of the \ac{UI}-part of \ac{LIMoSim} to enable live visualization of \ac{OMNeT++} key performance indicators.

%% file: tex/acknowledgment.tex
\section*{Acknowledgment}

\footnotesize
Part of the work on this paper has been supported by Deutsche Forschungsgemeinschaft (DFG) within the Collaborative Research Center SFB 876 ``Providing Information by Resource-Constrained Analysis'', project B4 and “European  Regional  Development  Fund”  (EFRE)  2014-2020 in  the  course  of  the  “InVerSiV”project  under  grant number EFRE-0800422 and has been conducted within the AutoMat (Automotive Big Data Marketplace for  Innovative  Cross-sectorial  Vehicle  Data  Services)  project, which received funding from the European Union’s Horizon 2020 (H2020) research and innovation programme under the Grant Agreement No 644657.